\documentclass[aps,prl,twocolumn,superscriptaddress]{revtex4}
\usepackage{graphicx}

\begin{document}

\title{Electromagnons in multiferroic YMn$_2$O$_5$ and TbMn$_2$O$_5$}

\author{A. B. Sushkov}
\affiliation{Materials Research Science and Engineering Center,
University of Maryland, College Park, Maryland 20742}

\author{R. Vald\'{e}s Aguilar}
\affiliation{Materials Research Science and Engineering Center,
University of Maryland, College Park, Maryland 20742}

\author{S. Park}
\affiliation{Department of Physics and Astronomy, Rutgers
University, Piscataway, New Jersey 08854}

\author{S-W. Cheong}
\affiliation{Department of Physics and Astronomy, Rutgers
University, Piscataway, New Jersey 08854}

\author{H. D. Drew}
\affiliation{Materials Research Science and Engineering Center,
University of Maryland, College Park, Maryland 20742}


\date{\today}

\begin{abstract}
Based on temperature dependent far infrared transmission spectra of 
YMn$_2$O$_5$ and TbMn$_2$O$_5$ single crystals, we report the observation of
electric dipole-active magnetic excitations, or electromagnons, in these multiferroics.  
Electromagnons are found to be directly responsible for the step-like anomaly of the 
static dielectric constant at the commensurate--incommensurate magnetic transition and 
 are the origin of the colossal magneto-dielectric effect reported in these multiferroics.  
\end{abstract}


\maketitle

Multiferroics materials, which exhibit simultaneous magnetic and ferroelectric order, 
have attracted much attention recently because of the interesting physics of systems with 
coupled multiple order parameters and because of their potential for cross electric and magnetic functionality.  
Thus electrically accessible magnetic memory and processing and vise versa are important possibilities.  
Striking intrinsic multiferroic cross-coupling effects such as magnetic 
field induced polarization switching and giant magnetocapacitance have been reported at cryogenic temperatures.  
Fundamental interest in multiferrocity also derives from the strong interplay between magnetic frustration, ferroelectric order, and 
fundamental symmetry issues in phase transformations that characterize these materials~\cite{Kimura-113,Hur-nature,Lawes-vanadate,Mostovoy-spiral,RC}.  

The strong coupling between the magnetic and lattice degrees of freedom can also lead to a complex excitation in solids, the electromagnon. 
This is the low frequency branch of the mixed magnon-phonon excitation, which has the character 
of a magnon with electric dipole coupling to electromagnetic radiation. 
It is responsible for dielectric anomalies at the phase boundaries and the colossal magneto-dielectric effect~\cite{Hur-Dy}.   
The possibility of electromagnon excitations in multiferroics has long been anticipated theoretically~\cite{Smol-Chupis}. 
The main effect of the magnon-phonon coupling is to produce shifts of their resonance frequencies (mode `repulsion') and a partial exchange of 
magnetic and electric dipole intensities.  
While dynamic spin-lattice coupling is present in all magnetic materials and leads to shifts of phonon 
frequencies (known as spin-phonon coupling) it is only in multiferroics that this coupling leads to detectable electromagnons.

The strongest coupling between magnetic and electric properties is found in improper magnetic ferroelectrics, 
such as TbMnO$_3$ \cite{Kimura-113}, TbMn$_2$O$_5$ \cite{Hur-nature}, and Ni$_3$V$_2$O$_8$ \cite{Lawes-vanadate}.  
A common feature of these compounds is frustrated magnetic interactions that result in non-collinear spin orderings.
Magnetoelastic coupling through the exchange interaction can produce improper ferroelectricity in systems with suitable symmetry.  
Within the underlying exchange interactions both symmetric and antisymmetric (or Dzyaloshinskii-Morya (DM)) type exchange can produce ferroelecticity.  
However, the DM exchange, due to its relativistic origin, is intrinsically weak so that symmetric exchange may 
be expected to be more favorable for strong coupling. 
Symmetry considerations are essential for unraveling the relevant interactions in these complex systems~\cite{RC}.  
While the direction of the spontaneous electric polarization~\textbf{P} is controlled by symmetry it does not 
uniquely identify the coupling mechanism~\cite{RC}. 
On the other hand, the selection rules for the electromagnons may be different for  
different exchange mechanisms, and therefore may help identify the dominant exchange processes involved in the multiferroicity. 

The role of the 3rd-order DM as well as 4th-order isotropic exchange coupling terms in dynamic coupling 
has been discussed~\cite{Smol-Chupis,Katsura-em}. 
It is understood that different coupling terms couple phonons to different magnon branches~\cite{Smol-Chupis}. 
In addition, recent model calculations show that electromagnons along the $b$-axis in YMn$_2$O$_5$ 
can be obtained from Heisenberg exchange and magnetostriction only~\cite{Mostovoy}.
          
Recently Pimenov et al. reported~\cite{Pimenov-Nature} the experimental observation of electromagnons in GdMnO$_3$ and TbMnO$_3$. 
However, this work leaves some ambiguity of the possible interference of the 
electromagnons with low energy transitions between crystal field split f-levels of the rare earth ions in these compounds. 
Golovenshits and Sanina~\cite{GS-JETPL} also reported mixed magnetic-lattice excitations in GdMn$_2$O$_5$ but 
no modes in EuMn$_2$O$_5$ were observed below their high frequency limit of 10~cm$^{-1}$. 
Therefore, this work is also ambiguous about the role of the rare earth f-electrons.

In this letter, we report on observations of electromagnons in the RMn$_2$O$_5$ (R = Y, Tb, Eu) multiferroics and compare them with previously reported electromagnons~\cite{Pimenov-Nature} as well as with `usual' magnons in multiferroic LuMnO$_3$.  
Our measurements on YMn$_2$O$_5$ eliminate any confusion, coming from the spin excitations in the rare earth ions. 
The polarization of the electromagnons can be additional evidence in favor of the symmetric exchange as the leading coupling mechanism.  

Single crystals of YMn$_2$O$_5$ and TbMn$_2$O$_5$ were grown as described elsewhere~\cite{Rolando}. 
The samples were characterized by X-rays and dielectric measurements in kHz range. 
Transmission measurements were performed using a Fourier-transform spectrometer in the frequency range from 4 to 200~cm$^{-1}$. 
The temperature dependence from 5 to 300~K was measured using liquid helium in a continuous flow cryostat (sample in vacuum) with optical access windows. 
We used a $^3$He-cooled bolometer for low frequency measurements. 
To increase the accuracy of transmission measurements, we prepared thicker samples for weak features and thin samples for strongest modes: 
0.46 and 0.09 mm for YMn$_2$O$_5$, and 0.725 and 0.16 mm for TbMn$_2$O$_5$.

In this work, we address three phases in TbMn$_2$O$_5$ (YMn$_2$O$_5$): LT1, incommensurate magnetic and ferroelectric, below 24(20) K; LT2, commensurate magnetic and ferroelectric, at 24(20)~$< T <$~38(41)~K; and the paramagnetic-paraelectric phase at $ T >$~41(45)~K. 
The low frequency transmission spectra of TbMn$_2$O$_5$ in three phases are shown in Fig.~1. 
The lowest phonon was found to be centered at 97~cm$^{-1}$ and is seen as zero transmission in Fig.~1b. 
We believe that all other features below this frequency are of magnetic origin (but with electric dipole activity). 
Figure~1a shows results for the strongest absorption (low transmission) near 10~cm$^{-1}$ in the LT1 phase in comparison to the paramagnetic phase. 
Figure~1b emphasizes weak absorption features. 

\begin{figure}
\includegraphics[width=\columnwidth]{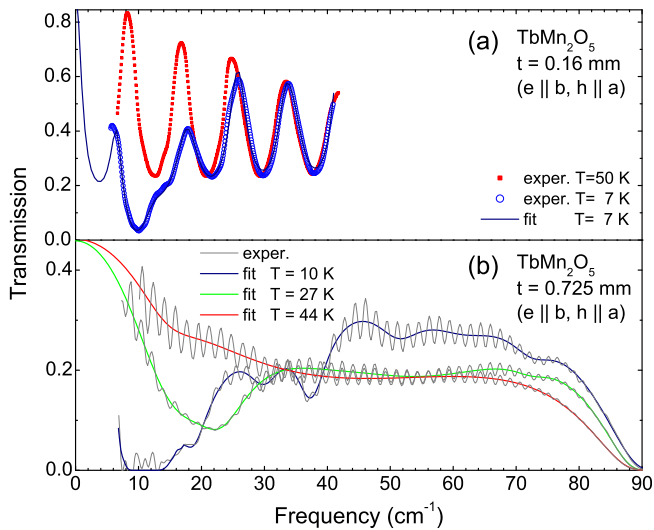}
\caption{(Color online) Transmission spectra of TbMn$_2$O$_5$: 
(a) thin sample, (b) thicker sample, oscillations are averaged out in model curves. 
Oscillations of experimental data are due to the etalon effect. $e$ and $h$ are electric and magnetic fields of light.
 } 
\label{Tb}
\end{figure}

Identifying these excitations as electromagnons requires addressing several questions.
To avoid confusion with possible transitions between f-levels of rare earth ions, we have studied YMn$_2$O$_5$.
A second issue is electric or magnetic dipole activity.  
Figure~2 presents transmission spectra of YMn$_2$O$_5$ for the three phases defined above. 
From fig.~2a it follows that the absorption bands are either $e || b$ or $h || a$ active. 
In the spectra in fig.~2b, measured for $h || a$ on a different cut of a crystal, there is no comparable absorptions. 
Spectra, taken in ($e || a$, $h || b$) configuration, are similar to those in fig.~2b --- no absorptions.  
It follows that strong absorptions occur only in $e || b ||$P orientation and are, therefore, electric dipole active. 
We do not exclude a possibility of observation of the magnetic activity in this frequency range, we just concentrate on the electric activity. 
Can these resonances be new phonons, activated in the low temperature phase? 
We have performed shell model calculations that put the lowest phonon near 100~cm$^{-1}$ for any reasonable parameters. 
Thus, we believe that we have definitively identified these features as electromagnons.

\begin{figure}
\includegraphics[width=\columnwidth]{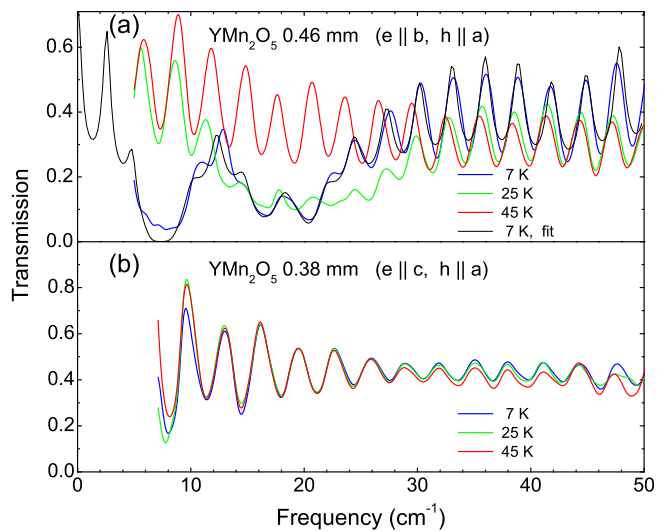}
\caption{(Color online) Transmission spectra of multiferroic YMn$_2$O$_5$: Absorption  
is observed only when $e || b$ and is, thus, electric dipole active. 
Nonzero transmission near 7~cm$^{-1}$ at 7~K in panel (a) and difference between curves in (b) is due to a low frequency noise. } 
\label{Y}
\end{figure}

To extract the parameters of the oscillators, we fit the transmission spectra 
with a Lorentzian model of the dielectric constant $\varepsilon(\omega)$ for electric dipole or magnetic permeability $\mu(\omega)$ 
for magnetic dipole transitions:
\begin{eqnarray}
	\varepsilon(\omega) = \varepsilon_\infty+\sum_j
\frac{S_j}{\omega_{j}^2-\omega^2-\imath\omega\gamma_j}
 \\
	\mu(\omega) = 1 + \sum_j
\frac{S_j}{\omega_{j}^2-\omega^2-\imath\omega\gamma_j}
\end{eqnarray}
where $\varepsilon_\infty $ is the high frequency dielectric constant, $j$ enumerates the oscillators,
$S_j$ is the spectral weight, $\omega_{j}$ is the resonance
frequency, and $\gamma_j$ is the damping rate. 
The extinction coefficient is proportional to $\sqrt{\varepsilon(\omega)\mu(\omega)}$. 
Thus, we can compare electric and magnetic dipole absorptions using same set of the oscillator parameters. 
These parameters are collected in the Table~1. 
Here, spectra of Y- and TbMn$_2$O$_5$ were fitted with formula (1) at fixed $\mu = 1$; and for LuMnO$_3$ we used (2) at fixed $\varepsilon=14.3$. 

\begin{table}[h]
\caption{Comparison of spectral parameters of electromagnons/magnons in various multiferroics.}
\begin{center}
\begin{tabular}{ l| c| c| c }
\hline
~        & $\omega_0$ & $S$ & $\gamma$  \\ 
Compound & (cm$^{-1}$) & (cm$^{-2}$) & (cm$^{-1}$)  \\
\hline
YMn$_2$O$_5$  &  7.2 &  170  & 1.9   \\
TbMn$_2$O$_5$ &  9.6 &  330  & 3.6   \\
TbMnO$_3$ \cite{Pimenov-Nature} &  23 &  $\approx$1000  & $>$25   \\
LuMnO$_3$     &  50 &  4  & 0.75   \\
\end{tabular}
\end{center}
\end{table}
 
It is interesting to compare the observed electromagnons to the uncoupled magnons in multiferroic hexagonal LuMnO$_3$. 
This compound is ferroelectric below $\sim$1000~K and antiferromagnetic below 76~K with geometric frustration. 
Using different cuts of the LuMnO$_3$ crystal, we observed one resonant absorption only in the $h(\omega) \perp c$ configuration. 
It means that this mode is magnetic dipole active (parameters are in Table~1). 
The dielectric constant shows a gradual `step-down' anomaly below 76~K, caused by phonon hardening~\cite{Souchkov}.  
Therefore, despite sharing with R125 the features of multiferroicity, frustrated magnetism, and non-collinear spin order, 
hexagonal LuMnO$_3$ does not have detectable electromagnon excitations. 
Hexagonal HoMnO$_3$ has several absorption bands below 100~cm$^{-1}$  which confirms the possible effects of rare earth ions. 

Figure 3 shows the optical conductivity of YMn$_2$O$_5$ and TbMn$_2$O$_5$ 
along the ferroelectric $b$-axis in three different phases as determined from the fits to the transmission spectra. 
In the ground state, LT1, both compounds show strong electromagnons at low frequencies and a several weaker peaks at higher frequencies. 
In the RMn$_2$O$_5$ structure 16 magnon branches are expected~\cite{Mostovoy}, and we see only strongest electromagnons in Fig.~3. 
In the LT2 phase only a broad mode, centered at 20~cm$^{-1}$, is present. 
This mode in Y has $S$=225~cm$^{-2}$ and contributes 0.6 to the static~$\varepsilon$. 
At 45~K, just above the Neel temperature, no well defined modes are observed. 
However, there is a broad background absorption which extends at least up to the lowest TO phonon frequency. 
This absorption is present at all three temperatures and it is really gone only at room temperature. 
We understand this broad absorption as frustrated spin fluctuations with electric dipole activity. 
The similar behavior of both compounds points to a leading role of Mn spin system in the physics of all of these electromagnon excitations. 

\begin{figure}
\includegraphics[width=\columnwidth]{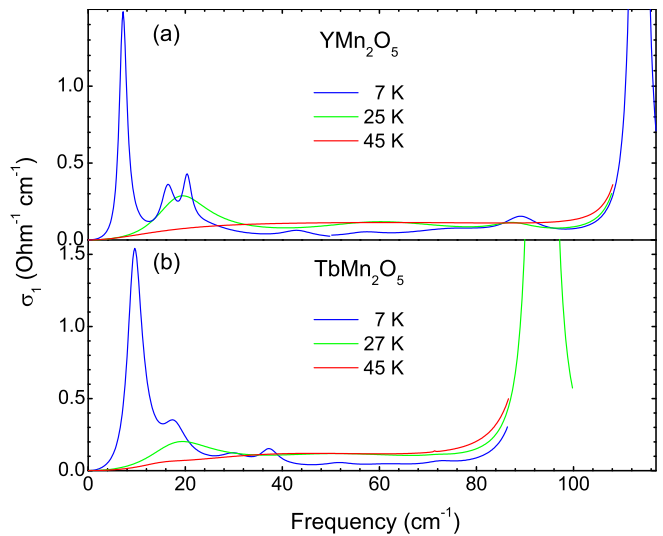}
\caption{(Color online) Optical conductivity of YMn$_2$O$_5$ and TbMn$_2$O$_5$ for $e || b$ in three phases. 
Strong peaks at 113 and 97~cm$^{-1}$ are the lowest phonons. Other peaks are electromagnons.  
} 
\label{s1}
\end{figure}

We also measured transmission of a EuMn$_2$O$_5$ polycrystal. Its absorption spectrum 
is similar to those shown in Fig.~3: a strong peak at 14~cm$^{-1}$ in the LT1 phase and a broad peak at 20~cm$^{-1}$ 
in the LT2 phase.  Therefore EuMn$_2$O$_5$ is another RMn$_2$O$_5$ compound with nonmagnetic rare earth ion that supports electromagnon excitations.

The data of Table~1 allow to compare the electromagnons in R113 and R125 compounds. 
The resonance frequencies of the strongest electromagnons in the R125 compounds are lower than those 
in TbMnO$_3$ suggesting a possible stronger mode `repulsion'.  However this is inconsistent with their weaker oscillator strength.
Therefore, the different frequencies may be due to the distinct magnetic anisotropies, which influence the bare magnon frequencies.
Electromagnon spectrum of the R125 compounds is better resolved into distinct modes than the R113 spectrum. 
However, the total spectral weight of $510$~cm$^{-2}$ of all TbMn$_2$O$_5$ modes, 
seen below the lowest phonon in LT1 phase, is closer to $\approx 1000$~cm$^{-2}$ of the TbMnO$_3$ electromagnon.
It would be interesting to compare infrared data to inelastic neutron scattering results 
to match electric and magnetic dipole activity in this frequency range for these two types of multiferroics. 

The polarization of electromagnons gives additional information about leading spin-lattice coupling mechanism. 
Katsura et al.~\cite{Katsura-em} showed that for DM type coupling and cycloidal spin chain the electromagnon 
is electrically active along the axis perpendicular both to the $k$-vector of spin structure and static polarization~\textbf{P}. 
These imply that the DM exchange is not the leading coupling in the RMn$_2$O$_5$ compounds where electromagnon is active for $e$$||$P. 
On the other hand, the latter polarization can be obtained from Heisenberg exchange and magnetostriction~\cite{Mostovoy}. 
Therefore, our results give additional evidence in favor of symmetric exchange being dominant in RMn$_2$O$_5$ 
compounds suggested earlier~\cite{Kadomtseva,Chapon-YMn2O5}.

It is important to know where the electromagnons borrow their electric activity.  
Simple model calculations emphasize the role of the lowest phonon~\cite{Katsura-em}. 
This prediction was qualitatively confirmed for GdMnO$_3$ by Pimenov et al.~\cite{PimenovCM} but the change in the phonon spectral weight (60~\%) was too large.
On examining the phonon spectra of TbMn$_2$O$_5$~\cite{Rolando}, we could not find any clear 
correlations with activation of electromagnons.  
The lowest phonon with $S\approx 5000$~cm$^{-2}$ is observed to strengthen slightly in the LT1 phase in contradiction to expectation from mode mixing. 
However, its frequency hardens in the LT phases~\cite{Rolando} in accord with the electromagnon models. 
Determining which phonons couple to the magnon and with what strength remains an important experimental challenge.
 
\begin{figure}
\includegraphics[width=\columnwidth]{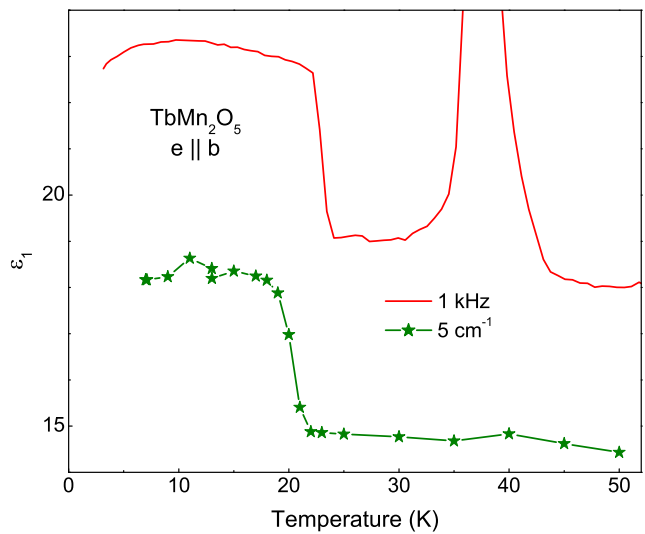}
\caption{(Color online) Dielectric constant of TbMn$_2$O$_5$ from fits of infrared spectra (lower curve) in comparison with kHz measurements.} 
\end{figure}

Figure 4 shows the temperature dependence of the static dielectric constant from kHz range measurements (top curve) compared with the dependence  
obtained from the fits of the far infrared spectra of TbMn$_2$O$_5$. 
We show only Tb compound data because of the larger sample size and higher frequency of the strongest electromagnon. 
These data show that the whole step-like anomaly in $\varepsilon(T)$ results from electromagnons. 
It follows that we observe practically all electromagnons. The largest 
contribution to $\varepsilon$ in the LT1 phase comes from the strong low frequency electromagnon.
The difference in absolute values of $\varepsilon$ comes from the zero-frequency ferroelectric mode. 
The sharp peak at 38~K (top curve), at the onset of \textbf{P}, is also produced by the zero-mode, 
and it is not observed in the far infrared (lower curve). 
The magnetic nature of electromagnons allows suppression/enhancement and/or Zeeman splitting of original magnon branches by externally applied 
magnetic fields~\cite{Pimenov-Nature,GS-JETPL} which leads to the magneto-dielectric effects~\cite{Hur-Dy}.

In conclusion, we report the observation of electromagnons in the RMn$_2$O$_5$ compounds including the non rare earth YMn$_2$O$_5$. 
The spectra of YMn$_2$O$_5$ and TbMn$_2$O$_5$ are very similar which proves the Mn origin of electromagnons in these compounds. 
The overall spectral weight of RMn$_2$O$_5$ electromagnons is approximately 1/2 that in RMnO$_3$ compounds. 
The electromagnons polarization selection rules in RMn$_2$O$_5$ provides evidence in favor of symmetric exchange coupling mechanism in this system. 
Additional far infrared and neutron measurements with a more complete theoretical analysis may provide a deeper understanding of multiferroicity.

\begin{acknowledgments}
We thank M.~Mostovoy and D.~Khomskii for useful discussions.  This work was
supported in part by the National Science Foundation MRSEC under
Grant No. DMR-0520471.
\end{acknowledgments}

\end{document}